\documentclass[11pt]{article}

 \advance \textwidth by -2in
 \advance \textheight by -3in
\begin{document}

\noindent {\bf TAKING NANOTECHNOLOGY TO SCHOOLS} \vskip 0.2cm

\noindent  {\bf Akhlesh Lakhtakia}
\vskip 0.2cm

\noindent {\sf CATMAS~---~Computational \& Theoretical Materials Sciences Group \\
\noindent Department of Engineering Science \& Mechanics\\
\noindent 212 Earth \& Engineering Sciences Building\\
\noindent Pennsylvania State University, University Park, PA
16802--6812} \vskip 0.4cm

\noindent {\bf ABSTRACT:} After a primer on nanotechnology and a review of current educational practices in secondary schools, the approach of just-in-time education is applied to integrate technosciences and humanities so that both future technoscientists and non-technoscientists develop a common understanding, possibly even a common language, to deal with social, ethical, legal, and political issues that arise from the development of nanotechnology and its convergence with other technoscientific developments.

\vskip 0.4cm

\noindent{\bf 1. INTRODUCTION}

Even during the 1980s, when its prospects could be faintly discerned by just a few researchers [1,2], nanotechnology promised a second industrial revolution. No wonder, total worldwide R\&D spending on nanotechnology in 2004 was an estimated USD 8,600 million [3] and continues to rise rapidly. The expenditure is remarkable, coming despite the fact that ``the 
relatively small number of applications $\dots$ that have made it through to industrial application represent evolutionary rather than revolutionary advances.'' [4]

Together with information technology, biotechnology, and cognitive science, nanotechnology is expected to radically alter the human condition within a short span of time probably not exceeding two decades. Human cultures however do not change at the same rapid pace~---~certainly not in times of relative peace~---~which renders it imperative that technoscientists as well as nonÐtechnoscientists begin to ponder and predict the parameters of possible social, legal and ethical changes to emerge in the first two decades of the new millennium. 

At the same time, the preparation of future leaders and an engaged citizenry to cope with those and other unpredicted changes must also begin in schools and colleges. In this article, the supplementation of current educational practices by a practice designed for that purpose is proposed. Section 2 presents a brief discussion of the many aspects of nanotechnology, Section 3 is a distillate of current educational practices in most countries, and Section 4 contains the proposed supplementary practice.

\vskip 0.4cm

\noindent{\bf 2. A NANOTECHNOLOGY PRIMER}

Nanotechnology is not a single process; neither does it involve a specific type of materials. Instead, the term ÒnanotechnologyÓ covers all aspects of the production of devices and systems by manipulating matter at the nanoscale.

Take an inch-long piece of thread; chop it into 25 pieces; and then chop one of those pieces into a million smaller pieces. Those itty-bitty pieces are about 1 nanometer (nm) long. The ability to manipulate matter and processes at the nanoscale undoubtedly exists in many academic and industrial laboratories. At least one relevant dimension must lie between 1 and 100 nm, according to the definition of nanoscale by the US National Research Council [5]. Ultrathin coatings have one nanoscale dimension, nanowires and nanotubes have two such dimensions, whereas all three dimensions of nanoparticles are at the nanoscale.  

Nanotechnology is being classified into three types. The industrial use of nanoparticles in automobile paints and cosmetics exemplifies {\em incremental\/} nanotechnology. Nanoscale sensors exploiting quantum dots and carbon nanotubes represent 
{\em evolutionary\/} nanotechnology, but their development is still in the embryonic stage. 
{\em Radical\/} nanotechnology, as envisioned in sci-fi thrillers such as Michael Crichton's 
{\em Prey,\/} does not seem viable in the next several decades.

Material properties at the nanoscale differ from those in bulk because of extremely large surface areas per unit volume at the nanoscale. Quantum effects also come into play at the nanoscale. Nanoscale properties and effects should transform current practices in integrated electronics, optoelectronics, and medicine. But the translation from the laboratory to mass manufacturing is currently fraught with significant challenges, and reliable manipulation of matter at the nanoscale in a desirable manner remains very difficult to implement economically.

Very little data exists on health hazards of nanotechnology. Because small amounts of nanomaterials are expected to be handled outside the workplace, a panel of experts convened by the Royal Society of London and the Royal Academy of Engineering [4] concluded that the risk to general public is minimal. However, the risk to nanoindustrial workers from inhalation as well as by skin penetration could be very high, and toxicological studies should be undertaken soon. There is also the risk of spontaneous combustion of nanomaterials due to the large surface-to-volume ratio.

Nanotechnology is emerging at a crucial stage of the global civilization. A remarkable convergence of nanotechnology, biotechnology, and information technology is occurring. Among the extremely pleasant prospects of their symbiosis are new medical treatments, both preventive and curative; monitoring systems for buildings, dams, ships, aircraft, and other structures vulnerable to natural calamities and terroristic acts; energy-efficient production systems that produce very little waste; and so on.

The convergence of the three technologies, along with progress in cognition science, also raises prospects that can be horrifying to contemplate. Consider the futuristic scenarios revealed in movies such as {\em Gattaca\/} and {\em The Manchurian Candidate.\/} Private watchdog groups must be formed to oversee governmental operations that can erode individual rights and privacy, while laws must be formulated to curb the powers of the controllers of the technologies over their users.

\vskip 0.4cm

\noindent{\bf 3. CURRENT EDUCATIONAL PRACTICES}

Secondary school curriculums in sciences and mathematics are primarily of two types. A student has to take two to three science courses and two to three mathematics courses very year for five or six years, in the first type of curriculums. This type is common in India and many Commonwealth countries, China, and many European countries. In the second type of curriculum, one science course and one mathematics course are taken every year for four to five years, the practice being very common in the USA.

The first type of curriculum is integrated across the scientific disciplines (physics, chemistry, biology, etc.) and mathematical disciplines (algebra, geometry, trigonometry, etc.) both horizontally and vertically. For instance, the mathematics required for science courses at a certain grade level was taught mostly in the previous grade level, and perhaps somewhat in the same grade level but earlier in the year. Likewise, word problems in mathematical courses may model scientific phenomenons taught earlier.

The second type of curriculum may have some horizontal integration but generally lacks vertical integration. In the US, earth sciences are taught in the 9th grade, biology in the 10th grade, chemistry in the 11th grade, and physics in the 12th grade. There is also a similar stratification of algebra, geometry, trigonometry, and calculus. This stratification may have once made sense to educationists, but is often derided by practising scientists, engineers and mathematicians. It is also inimical to the interdisciplinary and multidisciplinary mindset that must be inculcated for nanotechnology education at the late pre-university and university levels. Yet, the second type of curriculum does not burden the studentÕs mind unduly by rote learning and thus appears to promote innovation.

During the last two decades, the practices of collaborative learning and active learning have been incorporated in pre-university education to varying degrees. Collaborative learning practices involve the creation of student teams to carry out certain tasks, whether in laboratories or in classrooms or after school hours. This is valuable practice for future technoscientists, because much technoscientific activity is undertaken by teams of individuals with disparate areas of expertise. Active learning requires the supplementation of abstract principles by illustrative hands-on activities undertaken by students. These hands-on activities involve real-life equipment and situations as well as models mimicking certain aspects thereof, thus providing relevance to abstract ideas. Collaborative and active learning activities are often project-based: either individually or collectively, students have to complete various projects and present end-of-the-line deliverables to their fellow-students and teachers.

Current practices with both types of curriculums constitute just-in-case education (JICE). Instruction on certain topics is imparted just in case those topics turn out to be useful to the students in later years. There is much merit in JICE. The future cannot be predicted; hence, it is best that students acquire a broad background. Indeed, many technoscientists today attribute their success to 
JICE at the pre-university level. But successful education in nanotechnology~---~as also in information technology and biotechnology~---~would require supplemntation by a different instructional approach.

\vskip 0.4cm

\noindent{\bf 4. JUST-IN-TIME EDUCATION}

This new approach can be called called just-in-time education (JITE). The JITE approach was enunciated in 1992 [6], drawing upon the principles of JIT manufacturing. Today, it is heavily applied in the information technology and distance education sectors for training and retraining the adult workforce, as googling will readily prove. It is also used to enhance learning in heavily subscribed lower-division undergraduate courses at universities [7]. 

JITE appears ideally suited to address complex issues and to solve multidisciplinary problems,
in general. For taking nanotechnology to grade schools, the hallmark of this approach is that students shall have to learn: 
\begin{itemize}
\item
first, to identify the disciplines intersecting a complex problem; 
\item 
second, to acquire the necessary pieces of information and understanding from each intersecting discipline; 
\item
third, to synthesize the various parts into a whole that denotes an acceptable, if not desirable, level of accomplishment; 
\item 
fourth, to assess requirements for further developments; and 
\item 
fifth, to establish the values of their accomplishment in the cultures of their surroundings, nation, and the world. 
\end{itemize}

JITE is envisaged in terms of end-of-semester, end-of-school-year, and end-of-pre-university-education experiences. A JITE experience is a project that spans at least two  but preferably more scientific and mathematical disciplines. A project may be undertaken by a single student or a team of student, as appropriate, and every student must undertake single-member as well as team projects.

Organization and communication skills must be taught, even though some students may possess such skills. In a team project, individuals must be apportioned specific tasks whose completion must be reported to the team before certain deadlines. Tasks and reporting deadlines should also be delineated for single-member projects as well. Reporting can be either oral or written or both, as appropriate.

Crucially, only a part of the necessary information should have been imparted to the students in regular coursework prior to the commencement of the project. The remainder of the necessary information must be gathered from different sources~---~schoolbooks, extracurricular books, the web, site visits, and interviews with practitioners. Students must be encouraged to think of projects as not requiring standard, back-of-the-book answers; rather, different teams undertaking the same project could arrive at different conclusions and deliverables.

Introspection and reflection constitute another crucial aspect of JITE. The value of project tasks to the student must be assessed by him/her before and after undertaking each task. This is greatly facilitated by the student keeping a daily journal of activities as well as ideas. At the end of a project, every student must submit a statement of personal growth~---~what he/she had expected during the initial stages of the project, and what was actually learnt by the end of the project.

The statement of personal growth could incorporate nontraditional objectives. It could contain reflections on  (i) the relevance of the project to the town, province, nation, and the world; (ii) enhancement of cultural and ecological diversity and sustainability; and (iii) suggestions for follow-up projects and other activities. 

Not only would the practice of writing journals and statements of personal growth encourage healthy introspection, it would also anchor the scientific and mathematical disciplines with humanities and social sciences. This is especially important because the extraordinary convergence of nanotechnology, information technology, and biotechnology creates significant social, legal, political, and ethical issues that must be  effectively tackled by the citizenry of any country concurrently with technoscientific developments. 

Consequently, JITE is envisaged in the context of nanotechnology, etc., to not only impact the teaching and learning of science and mathematics but also of humanities and social sciences. JITE experiences shall have to be guided by teams of teachers drawn from a diverse array of disciplines encompassing language arts, sociology and history, civics and political science, physics, chemistry, biology, and mathematics. Science and mathematics teachers shall have to learn humanities and social sciences; but, even more importantly, humanities and social science teachers shall have to learn mathematics and sciences. Finally, only teachers who are themselves lifelong learners shall be able to effectively turn their students into lifelong learners. 

In some countries such as the US, JITE could entail prolongation of the school year, with attendant effects~---~very possibly deleterious~---~on vacation-related economic activities and higher school taxes. In other countries such as India and China, JICE activities would have to be reduced to accommodate JITE activities without increasing the schoolyear in length. In-service teachers would have to be trained in JITE, curriculums in colleges of education shall have to restructured, and teacher-certification procedures shall have to be revised.

\vskip 0.4cm

\noindent{\bf 5. CONCLUDING REMARKS}

The relentless development of major new technologies that are continually transforming the social, legal, and ethical parameters of human societies strongly indicates that current educational practices should be altered in order to produce future leaders as well as an informed and engaged citizenry. The proposed supplementation of current educational practices by just-in-time education experiences will integrate the humanities with technosciences, hopefully seamlessly, thereby instilling 
\begin{itemize}
\item[(i)]	in future social, political, administrative and judicial leaders the abilities to cope with and perhaps anticipate changes wrought by technoscientific developments;
\item[(ii)]	in future scientific and technological leaders the abilities to forecast the socioethical values of their developments; and
\item[(iii)]	in all citizens the abilities to assess the significance of techoscientific developments, and thus be engaged in opinion-making forums and decision-making legislatures.
\end{itemize}

\vskip 0.4 cm
\noindent{\bf Acknowledgements.} The author gratefully acknowledges several discussions with Drs. Debashish Munshi and Priya Kurian (University of Waikato, Hamilton, New Zealand), Fei Wang (Micron Corp., Boise, ID, USA), Russell Messier and Jeffrey M. Catchmark (Pennsylvania State University), and Arnold Burger (Fisk University, Nashville, TN, USA). Thanks are also due to Joseph B. Geddes III for feedback on an early draft.

\vskip 0.4 cm
\noindent {\bf References}

\begin{enumerate}

\item
K. E. Drexler, Molecular engineering: an approach to the development of general capabilities for molecular manipulation, {\em Proc. Nat. Acad. Sci.\/} {\bf  78}, 5275-5278 (1981).
\item
K. E. Drexler, 
{\em Engines of creation: The coming era of nanotechnology\/} (New York: Anchor-Doubleday, 1986).
\item
{\em The Economist\/}  (2005, Jan 1-7 issue).
\item
Royal Society and Royal Academy of Engineering, {\em Nanoscience and nanotechnologies: Opportunities and uncertainties\/}, RS Policy Document 19/04 (London: Royal Society, 2004). Also available at {\sf http://www.nanotec.org.uk/finalReport.htm}.
\item
National Research Council, {\em Condensed-matter and materials physics: Basic research for tomorrow's technology\/} (Washington, DC: National Academy Press, 1999).
\item
D. Hudspeth, Just-in-time education, {\em Educ. Technol.\/}, 7-11 (1992, June issue).
\item
G. M. Novak, E. T. Patterson, A. Gavrin and R. C. Enger,  Just-in-time teaching: Active learner pedagogy with www (1998), {\sf  http://webphysics.iupui.edu/JITT/ccjitt.html} (1998).
\end{enumerate}

\end{document}